\documentclass[prb,twocolumn,showpacs,preprintnumbers,amsmath,amssymb,supersciptaddress]{revtex4}
\usepackage{graphicx}
\usepackage{verbatim}

\begin{document}

\title{Comparing $GW$+DMFT and LDA+DMFT for the testbed material  SrVO$_3$}

\author{C. Taranto$^{1}$}
\author{M. Kaltak$^{2}$}
\author{N. Parragh$^{1,3}$}

\author{G. Sangiovanni$^{1,3}$}

\author{G. Kresse$^{2}$}

\author{A. Toschi$^{1}$}

\author{K. Held$^{1}$}

\affiliation{$^1$Institute for Solid State Physics, Vienna University of Technology, 1040 Vienna, Austria}
\affiliation{$^2$University of Vienna, Faculty of Physics and Center for Computational Materials Science, Sensengasse 8/12,
A-1090 Vienna, Austria}
\affiliation{$^3$Institut f\"ur Theoretische Physik und Astrophysik,
  Universit\"at W\"urzburg, Am Hubland, D-97074 W\"urzburg, Germany}
\date{Version 0, \today }

\begin{abstract}
We have implemented the $GW$+dynamical mean field theory (DMFT) approach in the Vienna ab initio simulation package.  Employing the interaction values obtained from the locally unscreened random phase approximation (RPA), we compare $GW$+DMFT and LDA+DMFT against each other and against experiment for SrVO$_3$. We observed a  partial compensation of stronger electronic correlations due to the reduced $GW$ bandwidth and weaker correlations due
to a larger screening of the RPA interaction, so that the obtained spectra are quite similar and well agree with experiment. Noteworthily, the $GW$+DMFT better reproduces the position of the lower Hubbard side band.
\end{abstract}


\maketitle

\let\n=\nu \let\o =\omega \let\s=\sigma

\section{Introduction}

The
local density approximation (LDA) plus dynamical mean field theory (DMFT)
approach  \cite{P7:Anisimov97a,P7:Lichtenstein98a,P7:LDA+DMFT1,P7:LDA+DMFT2,P7:LDA+DMFT3} has been
 a significant step forward for calculating materials with strong electronic correlations. This is, because--- on top of the LDA ---DMFT\cite{DMFT,DMFTREV} includes a major part
of the electronic
correlations: the local ones.
In recent years,
 LDA+DMFT has been
 applied successfully to many materials and correlated electron phenomena, ranging from transition metals and their oxides to rare earth and their alloys,
for reviews see Refs.\ \onlinecite{P7:LDA+DMFT2,P7:LDA+DMFT3}.

For truly parameter-free {\em ab initio} calculations, however, two severe  shortcomings persist: (i)  
the screened Coulomb interaction is usually considered to be an
adjustable parameter in LDA+DMFT
and (ii) the so-called double counting problem  (i.e., it is difficult
to determine  the electronic correlations already accounted for
at the LDA level). 
These shortcomings are  intimately connected with the fact that the 
density functional nature  of LDA does not match. 
with the many-body, Feynman-diagram structures of DMFT. Hence, it is not clear
what correlations are already included on the LDA level
or how to express these as a self-energy to avoid a double
counting with DMFT correlations.
These problems can be mitigated, but not solved,   
by constrained LDA  (cLDA) calculations  \cite{Dederichs84,McMahan88,Gunnarsson89}, which can 
be exploited to extract two independent parameters: interaction and double
counting correction. \cite{McMahan88,JCAMD}

A conceptionally preferable and better defined many-body  approach is achieved if one substitutes LDA by the so-called  $GW$ approximation  \cite{P7:GW,P7:Aryasetiawan98}.
 Since its proposition   by Biermann {\em et al.} \cite{P7:GW+DMFT}, 
the development and application of such a 
$GW$+DMFT scheme for actual applications has been tedious. This is reflected in the number of LDA+DMFT calculations for actual materials, which is of the order of a few hundred, compared to
a single\cite{note_jan} $GW$+DMFT calculation for Ni \cite{P7:GW+DMFT},
despite many advantages of $GW$+DMFT, such as the possible rigorous definition in terms of Feynman diagrams  and the avoidance of introducing {\sl ad hoc} parameters for the Coulomb interaction and double counting corrections.
 The reason for this imbalance  is twofold. First, since
the $GW$ approach 
is computationally fairly demanding and complex, 
mature $GW$ programs were missing in the past. Second,
the $GW$+DMFT scheme is considerably more involved than LDA+DMFT, in particular,
if calculations are done self-consistently and with a frequency dependent
(screened) Coulomb interaction. 
Indeed, these concepts are presently tested on the model
level \cite{Sun02,P7:Aryal12a}. Let us also note in this context that a frequency dependent
 interaction has  been employed on top of an LDA bandstructure
for BaFe$_2$As$_2$ \cite{P7:Werner12} and SrVO$_3$  \cite{P7:Casula12b}

In this paper, we present  results of our  $GW$+DMFT implementation in the 
Vienna ab initio simulation package
(VASP)
 for  SrVO$_3$ and compare the $GW$+DMFT results to those of LDA+DMFT \cite{DMFTSrVO3} as well as photoemission spectroscopy\cite{SrVO3exp}. We find the  $GW$+DMFT spectra to be quite similar
to that of LDA+DMFT due to a partial cancellation of two effects: the reduced 
$GW$ bandwidth in comparison to LDA and the weaker screened Coulomb interaction. An important difference, however, is the position of the lower Hubbard band
which in $GW$+DMFT better agrees with experiment.
To mimic the frequency dependence of the Coulomb interaction, which we have not included, we 
also performed $GW$+DMFT calculations for a  Bose factor ansatz \cite{P7:Casula12}
reduced bandwidth. The obtained spectra 
are rather different from $GW$+DMFT without Bose factor and LDA+DMFT.

 The paper is organized as follows: In Section \ref{Sec:method}, we discuss the
method and implementation. In Section  \ref{Sec:results}, we compare
LDA+DMFT and $GW$+DMFT self energies and spectral functions.
 A comparison to photoemission  experiments is provided in
Section  \ref{Sec:experiment}
and a conclusion in Section  \ref{Sec:conclusion}.

\section{Method}\label{Sec:method}
Let us briefly outline the relevant methodological aspects. 
Starting point of our calculation is the  $GW$ implementation within
(VASP). \cite{P7:Shishkin06}
Specifically, we first performed Kohn Sham density functional theory calculations 
using the local density approximation for SrVO$_3$ at the
LDA lattice constant of $a=3.78~${\AA}  and
determined the Kohn Sham  one-electron orbitals $\phi_{n\mathbf{k}}$ and one-electron energies $\epsilon_{n\mathbf{k}}$.
The position of the GW quasiparticle peaks were calculated by 
solving the linear  equation
\begin{equation}
\label{equ:qpe}
\begin{split}
 E_{n\mathbf{k}}^{QP} & = \epsilon_{n\mathbf{k}}+Z_{n\mathbf{k}}\times   \\
&  Re[\langle \phi_{n\mathbf{k}}|T+V_{n-e}+V_{H} + \Sigma(\epsilon_{n\mathbf{k}}) | \phi_{n\mathbf{k}}\rangle-  
\epsilon_{n\mathbf{k}}],
\end{split} 
\end{equation}
where $T$ is the one-electron kinetic energy operator and $V_{n-e}$  and $V_{H}$
are the nuclear-electron potential and the Hartree-potential, respectively. $\Sigma$
is the $G_0W_0$ self energy, and $Z_{n\mathbf{k}}$ is the renormalization factor evaluated
at the Kohn-Sham eigenvalues.\cite{Louie,P7:Shishkin06}
The original Kohn Sham orbitals are maintained at this step.
The Kohn Sham orbitals expressed in the projector augmented wave (PAW) basis are then projected onto
maximally localized Wannier functions 
\cite{P7:Marzari97} using the Wannier90 code.\cite{Wannier90}
To construct an effective low-energy Hamiltonian for the $t_{2g}$ vanadium orbitals,
we follow  Faleev, van Schilfgaarde and Kotani and approximate
the frequency dependent $G_0W_0$ self-energy by an  Hermitian operator $ \bar H$
that reproduces the position of the quasiparticle peaks of the original self-energy exactly:\cite{P7:Faleev04,P7:Chantis06}
\begin{equation}
 \bar H_{mn,\mathbf{k}} = \frac{1}{2}[\langle \phi_{m\mathbf{k}} | \Sigma^*(E_{m\mathbf{k}}^{QP})+\Sigma(E_{n\mathbf{k}}^{QP}) | \phi_{n\mathbf{k}} \rangle].
\end{equation}
In practice, for the present calculations, we have applied the slightly more involved
procedure to derive an Hermitian  approximation  
outlined in Ref. \onlinecite{Shishkin07}, although this yields
essentially an identical Hermitian operator $\bar H_{mn,\mathbf{k}}$.
Furthermore the off diagonal components are found to be negligible small, and henceforth disregarded.
The final Hermitian and $\mathbf{k}$-point dependent operator $\bar H$ is transformed to the Wannier
basis and passed on to the DMFT code, where it is used to construct the
$\mathbf{k}$-dependent self energy by adding the local DMFT self energy. 

\begin{figure}[tb]
\includegraphics[width=7cm]{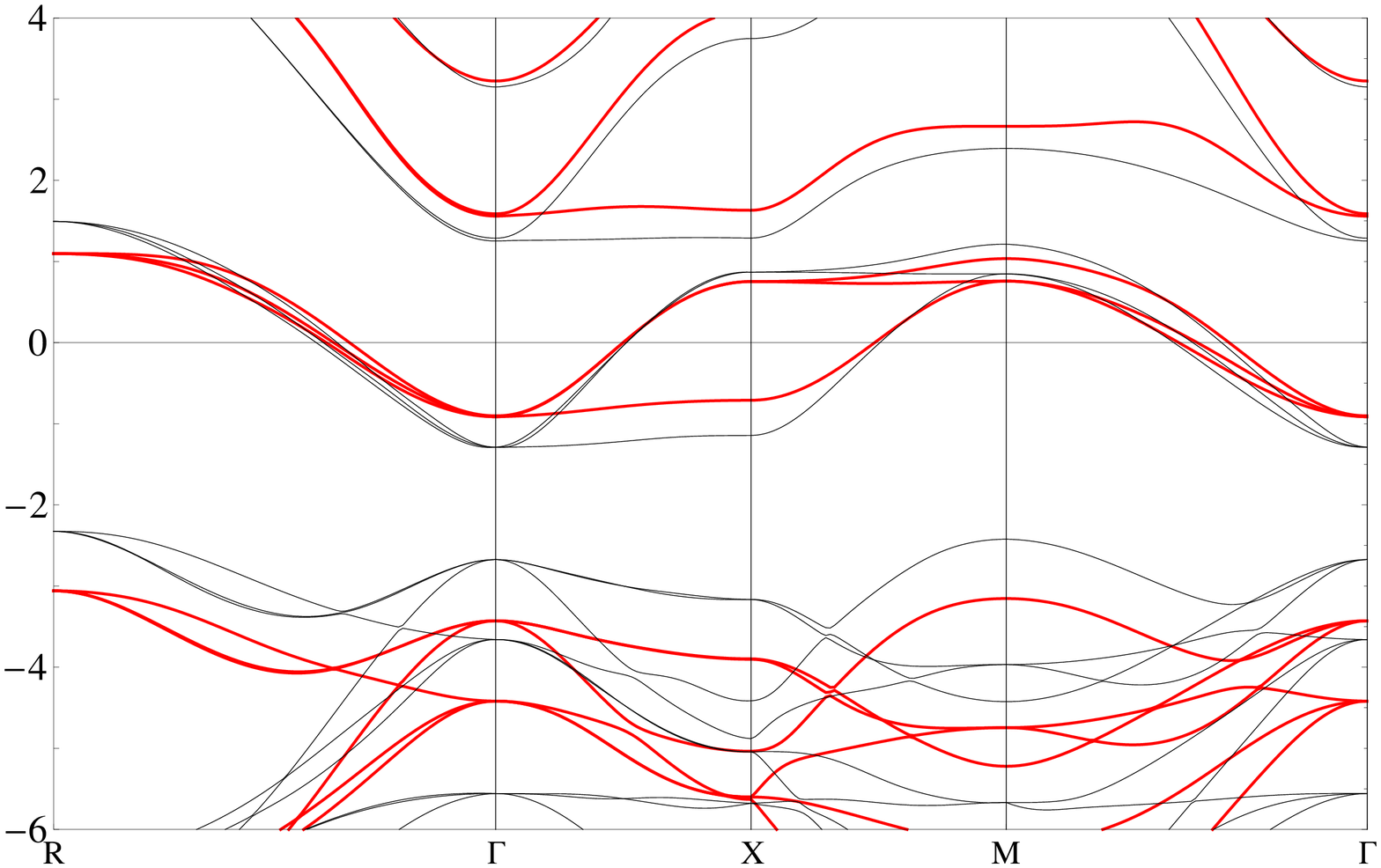}
\includegraphics[width=7cm]{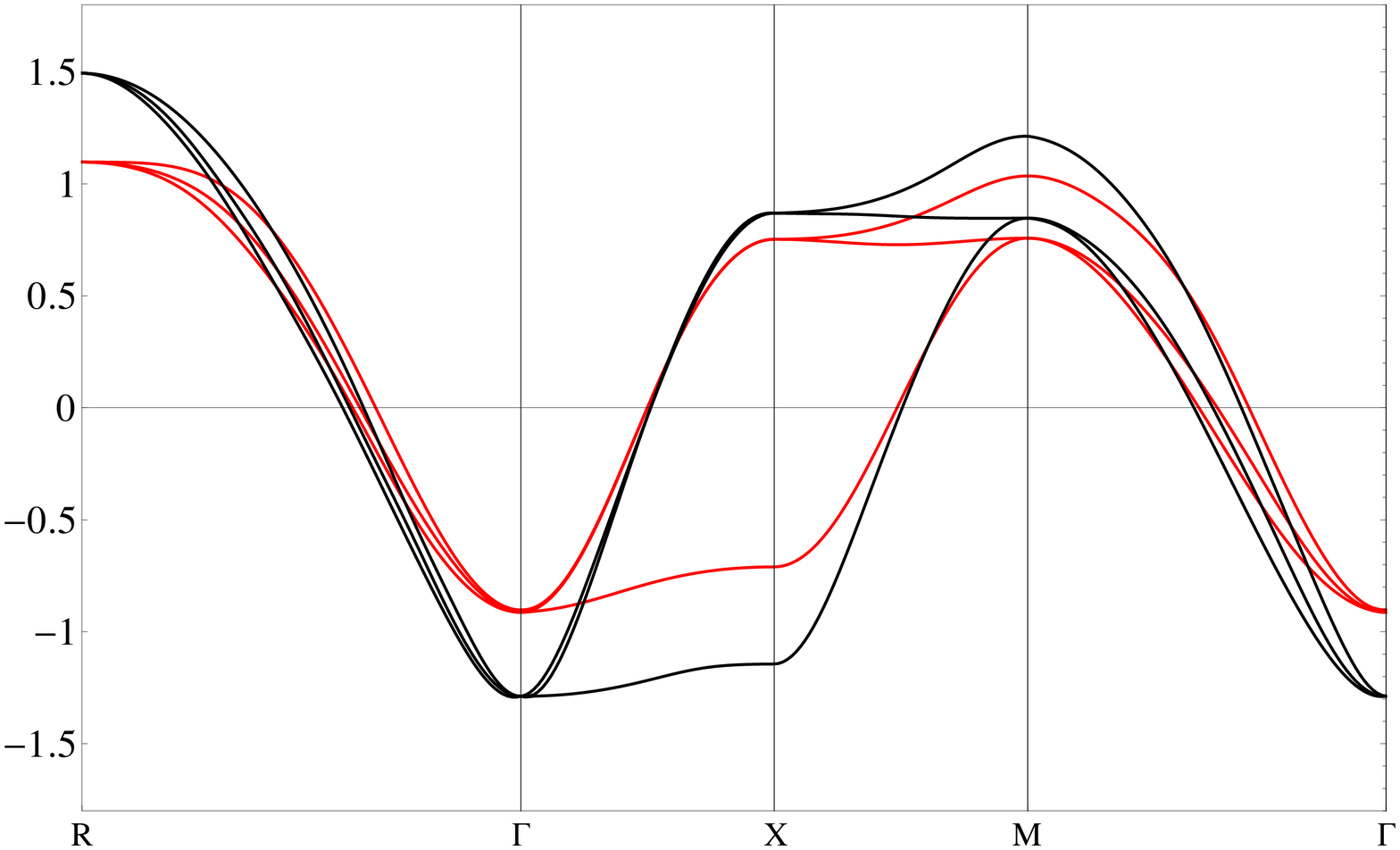}
\caption{(Color online) Upper panel: $G_0W_0$ quasiparticle bands (red, grey)  in comparison to LDA (black). The Fermi level sets our zero of energy and is marked as a line. Lower panel: Wannier projected $t_{2g}$ bandstructure from  $G_0W_0$ (red, grey) and LDA (black). The $t_{2g}$ target bands bandwidth is reduced by $\sim 0.7\,$eV in $GW$.}
\label{wannier_bands}\label{wannierbands}
\end{figure}

Fig.\ \ref{wannierbands} shows the obtained $G_0W_0$ bandstructure, which
for  the $t_{2g}$ vanadium target bands is about 0.7\,eV narrower than for the
LDA. The oxygen $p$ band (below $-2$\,eV) is shifted downwards by 0.5\,eV compared to the LDA,
whereas the vanadium $e_g$ bands (located about 1.5\,eV above the Fermi-level)
are slightly shifted upwards by 0.2\,eV. In the LDA, the top most vanadium $t_{2g}$ band at the
$M$ point is slightly above the lowest $e_{g}$ band at the $\Gamma$ point, 
whereas the $G_0W_0$ correction opens a gap between
the $t_{2g}$ and  $e_{g}$ states.

For brevity we will refer to the  subsequent DMFT calculations
as GW+DMFT, although it should be kept in mind that no self-consistency
is performed in the $GW$ part and furthermore the $GW$ self energy is approximated
by an Hermitian (frequency in-dependent) non-diagonal operator. Subtracting the local part of this Hemritian operator (to avoid a double counting) yields for the degenerate $t_{2g}$ orbitals a constant shift. This procedure
allows to maintain the structure and outline of the common DFT-DMFT scheme
and can be easily adopted in any DMFT code. However it
neglects lifetime broadening introduced by other bands not treated
on the DMFT level.

Within this Wannier basis, we  also calculate
the screened Coulomb interaction using the random phase approximation (RPA).
As described in  Ref.\  \onlinecite{P7:Nomura12} for an accurate estimate of the interaction value to be used in DMFT ($U^{\rm DMFT}$), only the local screening processes of the
 $t_{2g}$ target bands of  SrVO$_3$ are disregarded since  only
these are considered later on in DMFT. This approach \cite{P7:Nomura12}
is similar to the constrained RPA (cRPA) \cite{P7:Aryasetiawan04,P7:cRPA2}, with the difference being that in cRPA also {\em non-local} screening processes
of the  $t_{2g}$ target bands are disregarded which are not included in
DMFT. 
We carefully compare  $GW$+DMFT with LDA+DMFT calculations and experiment.
In both cases we use (frequency-independent) 
interactions obtained from this locally unscreened RPA and cLDA.
The Kanamori interaction parameters as 
derived from the locally unscreened RPA are: intra-orbital Coulomb repulsion
$U^{\rm DMFT}=3.44\,$eV; inter-orbital  Coulomb repulsion $\bar{U}^{\rm DMFT}=2.49\,$eV; Hund's exchange and pair hopping amplitude 
$J^{\rm DMFT}=0.46\,$eV. \cite{NoteUDMFT}
These values are, for SrVO$_3$, almost identical to the cRPA. \cite{P7:Nomura12}
In cLDA, on the other hand, somewhat larger interaction parameters were obtained and are employed by us for the corresponding calculations:
$U^{\rm cLDA}=5.05\,$eV,  $\bar{U}^{\rm cLDA}=3.55\,$eV,
$J^{\rm cLDA}=0.75\,$eV. \cite{SrVO3exp}

For the subsequent DMFT calculation, we employ the W\"urzburg-Wien w2dynamics code \cite{P7:Parragh12}, based on the hybridization-expansion
variant \cite{Werner05a} of the
continuous-time quantum Monte Carlo method (CT-QMC) \cite{Rubtsov04a}. This algorithm is particularly fast since
it employs additional  quantum numbers for
a rotationally-invariant Kanamori interaction 
 \cite{P7:Parragh12}. 
The maximum entropy method is employed for the
analytical continuation of the imaginary time and (Matsubara) frequency CT-QMC data 
to real frequencies
\cite{MEM}.

All our  calculations are without self-consistency, which is to some extend justified
for SrVO$_3$. Since the three $t_{2g}$ bands
of  SrVO$_3$ are degenerate, DMFT does not change the charge density of the low-energy $t_{2g}$ manifold and
hence self-consistency effects are expected to be small for LDA+DMFT.
This is, in principle, different for 
$GW$+DMFT. Here, the frequency dependence of the 
DMFT self energy might  yield  some feedback already for a simplified
Faleev, van Schilfgaarde and Kotani  quasiparticle self-consistency  \cite{P7:Faleev04,P7:Chantis06}.
Finally, we also test the Bose factor ansatz \cite{P7:Casula12} which renormalizes the $GW$ bandwidth by 
a renormalization factor ${\cal Z}_B\!=\!0.7$ \cite{P7:Casula12} for mimicking  the frequency dependence of the (partially unscreened) RPA interaction. 

\section{Results}\label{Sec:results}
\begin{figure}[tb]
\includegraphics[width=6cm]{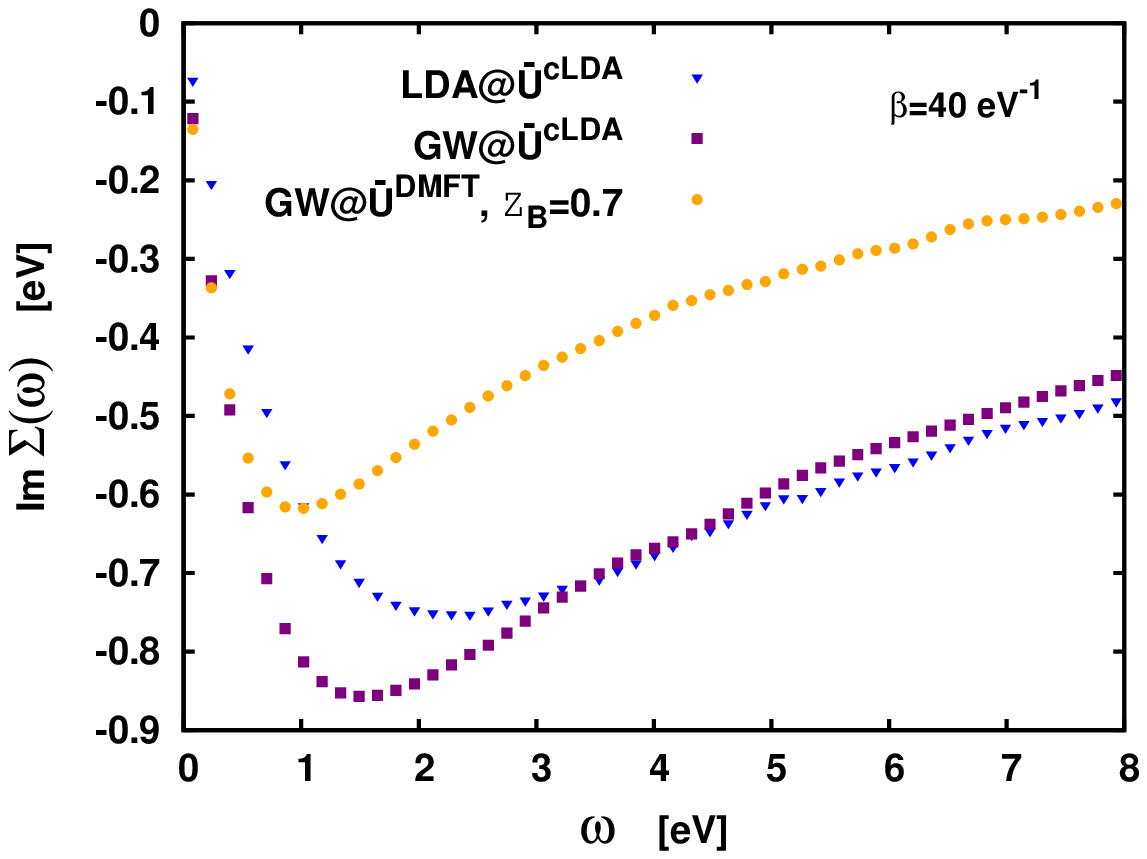}
\includegraphics[width=6cm]{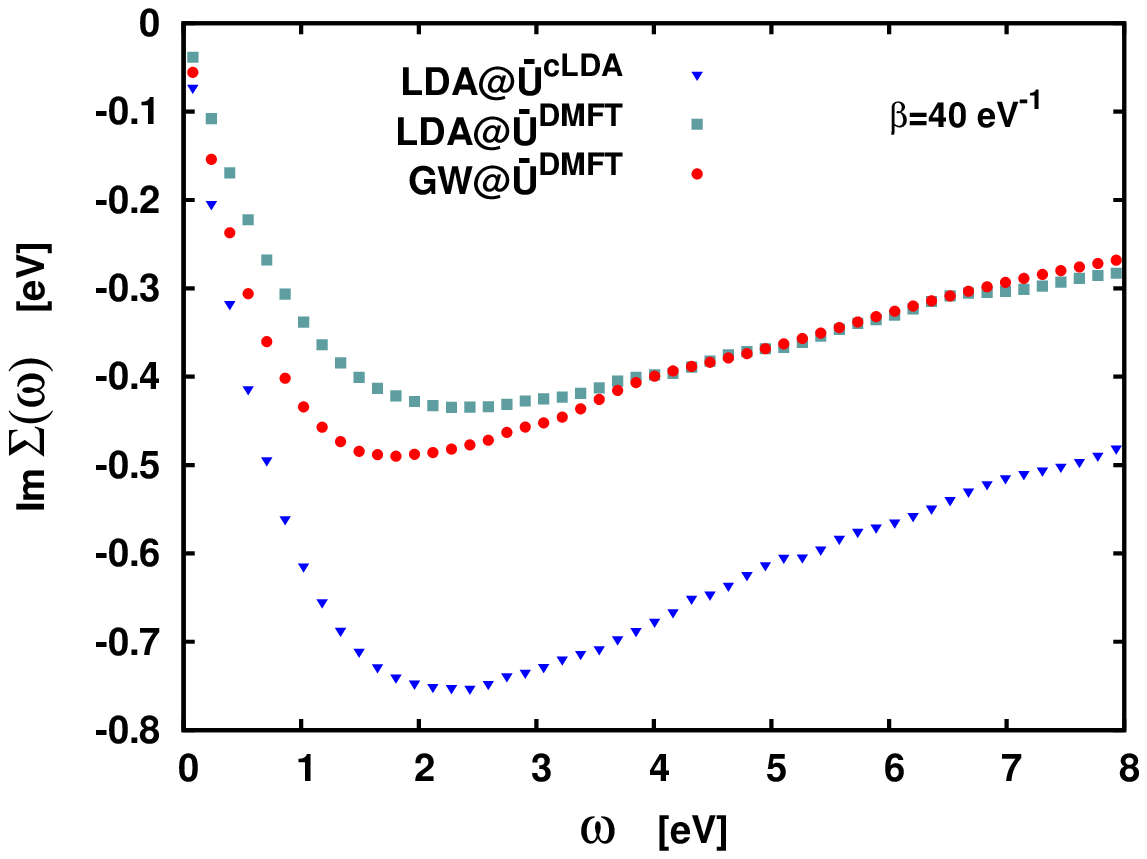}
\caption{(Color online) Comparison of the imaginary part of the DMFT self energies $\Sigma$ vs.\  (Matsubara) frequency $\omega$ for SrVO$_3$ at inverse temperature $\beta=40$ eV$^{-1}$ as computed in five different ways:
employing $GW$ and  LDA Wannier bands, the locally unscreened RPA interaction $\bar{U}^{\rm DMFT}=2.49$\, eV and the cLDA  $\bar{U}^{\rm cLDA}=3.55\,$eV, as well as the Bose factor renormalization
of ${\cal Z}_B\!=\!0.7$. \cite{P7:Casula12}}
\label{senergy}
\end{figure}

For analyzing the differences between $GW$+DMFT and LDA+DMFT we analyze and compare in the following five different calculations:
\begin{enumerate}
\item LDA+DMFT$@\bar{U}^{\rm cLDA}$ (conventional LDA+DMFT calculation with the cLDA interaction $\bar{U}^{\rm cLDA}=3.55\,$eV).
\item LDA+DMFT$@\bar{U}^{\rm DMFT}$ (LDA+DMFT calculation but with the locally unscreened RPA interaction $\bar{U}^{\rm DMFT}=2.49\,$eV).
\item $GW$+DMFT$@\bar{U}^{\rm DMFT}$ ($GW$+DMFT calculation with $\bar{U}^{\rm DMFT}=2.49\,$eV) 
\item $GW$+DMFT$@\bar{U}^{\rm cLDA}$ ($GW$+DMFT calculation but with $\bar{U}^{\rm cLDA}=3.55\,$eV) 
\item $GW$+DMFT$@\bar{U}^{\rm DMFT}$,${\cal Z}_B\!=\!0.7$ (as 3. but with a Bose renormalization factor ${\cal Z}_B$) 
\end{enumerate}
Let us first turn to the imaginary part of the local self energy which is shown
as a function of  (Matsubara) frequency in Fig. \ref{senergy}. The self energy yields a first impression
how strong the electronic correlations are in the various calculations.
The  LDA+DMFT$@\bar{U}^{\rm DMFT}$ self energy is the least correlated one, somewhat less correlated than LDA+DMFT$@\bar{U}^{\rm cLDA}$
due to the smaller locally unscreened Coulomb interaction  ($\bar{U}^{\rm DMFT}=2.49\,$eV $<$ 3.55\,eV $=\bar{U}^{\rm cLDA}$). 
For the same reason also the $GW$+DMFT$@\bar{U}^{\rm DMFT}$ self energy is less
correlated than that of a  $GW$+DMFT$@\bar{U}^{\rm cLDA}$ calculation.

If we compare LDA+DMFT and $GW$+DMFT on the other hand,
the  LDA+DMFT self energy is less correlated than
  the $GW$+DMFT one, if the Coulomb interaction is kept the same.
This is due to the 0.7\,eV smaller $GW$ $t_{2g}$-bandwidth
in comparison to LDA.  This  observation 
also reflects in the DMFT quasiparticle renormalization factors $Z$ 
which were obtained from a fourth-order fit to the lowest
four Matsubara frequencies, see  Table \ref{table1}.
Note that there is   an additional $GW$ renormalization factor
reducing the bandwidth in comparison to LDA.

However, the effect of the smaller $GW$ bandwidth partially
compensates with the smaller $\bar{U}^{\rm DMFT}$ interaction
strength. Altogether this yields 
rather  similar self energies 
of the standard approaches:   LDA+DMFT$@\bar{U}^{\rm cLDA}$ and  $GW$+DMFT$@\bar{U}^{\rm DMFT}$, see  lower panel of Fig. \ref{senergy}.
This also reflects in very similar renormalization factors in  Table \ref{table1}, $Z=0.51$ vs. $Z=0.57$, which both agree well  with experimental estimates
of 0.5-0.6 . \cite{SrVO3exp,Zexp}

\begin{table}
\begin{tabular}{l | r|r|r|r}
Scheme & $Z$& $d_{\rm intra}$& $d_{\rm inter}^{\uparrow\uparrow}$&$d_{\rm inter}^{\uparrow\downarrow}$ \\
\hline
LDA+DMFT$@\bar{U}^{\rm cLDA}$ &  0.51  &  0.004            &       0.013 &
0.009                \\
LDA+DMFT$@\bar{U}^{\rm DMFT}$ & 0.67  & 0.007        &           0.016 &
0.013 \\
$GW$+DMFT$@\bar{U}^{\rm DMFT}$ &  0.57  & 0.005             &      0.014 &
0.010  \\
$GW$+DMFT$@\bar{U}^{\rm cLDA}$ & 0.39   & 0.003   &                0.010 &
0.007 \\
$GW$+DMFT$@\bar{U}^{\rm DMFT}\!$,${\cal Z}_B\!=\!0.7$ \, &  0.36 &  0.003       &            0.009 & 0.006\\
experiment \protect \cite{SrVO3exp,Zexp} &$\sim$0.5-0.6 &&&\\
\end{tabular}
\caption{(Color online) DMFT quasiparticle renormalization factors  $Z$ from   the five different calculations at inverse temperature $\beta=40$\,eV$^{-1}$. Also shown are the pairwise double occupations within the same orbital $d_{\rm intra}$ and between different orbitals with the same $d_{\rm inter}^{\uparrow\uparrow}$ and opposite spin $d_{\rm inter}^{\uparrow\downarrow}$.  The ``standard'' LDA+DMFT$@\bar{U}^{\rm cLDA}$ and $GW$+DMFT$@\bar{U}^{\rm DMFT}$ calculations are similarly correlated and well agree with experiment. 
Using the cLDA interaction($\bar{U}^{\rm cLDA}$) for $GW$+DMFT or the locally unscreened RPA ($\bar{U}^{\rm DMFT}$) for LDA+DMFT yields a too strongly and too weakly correlated solution in comparison to experiment, respectively. Note that $GW$+DMFT  becomes even more strongly correlated, if the Bose renormalization factor is included.
 }
\label{table1}
\end{table}
Since one important difference is the strength of the interaction,
it is worthwhile recalling that $\bar{U}^{\rm DMFT}$ is defined as the local interaction strength at
low frequencies. While this value is almost constant within the range of
the $t_{2g}$-bandwidth, it is much larger at larger energies, exceeding $10$\,eV.
It has been recently argued and  shown in model
calculations \cite{P7:Casula12} that the stronger frequency-dependence of the screened Coulomb interaction at high energies
is of relevance and can be mimicked by a Bose factor renormalization of the
$GW$ bandwidth. The latter has been determined
as ${\cal Z}_B\!=\!0.7$ for SrVO$_3$. We have tried to take this into account
in the  $GW$+DMFT$@\bar{U}^{\rm DMFT}$,${\cal Z}_B\!=\!0.7$ calculation.
Due to the additional bandwidth renormalization, this calculation
is very different from all others and yields the largest quasiparticle
renormalization, i.e., $Z=0.36$ is smallest.

\begin{figure*}
\begin{center}
\includegraphics[width=16cm]{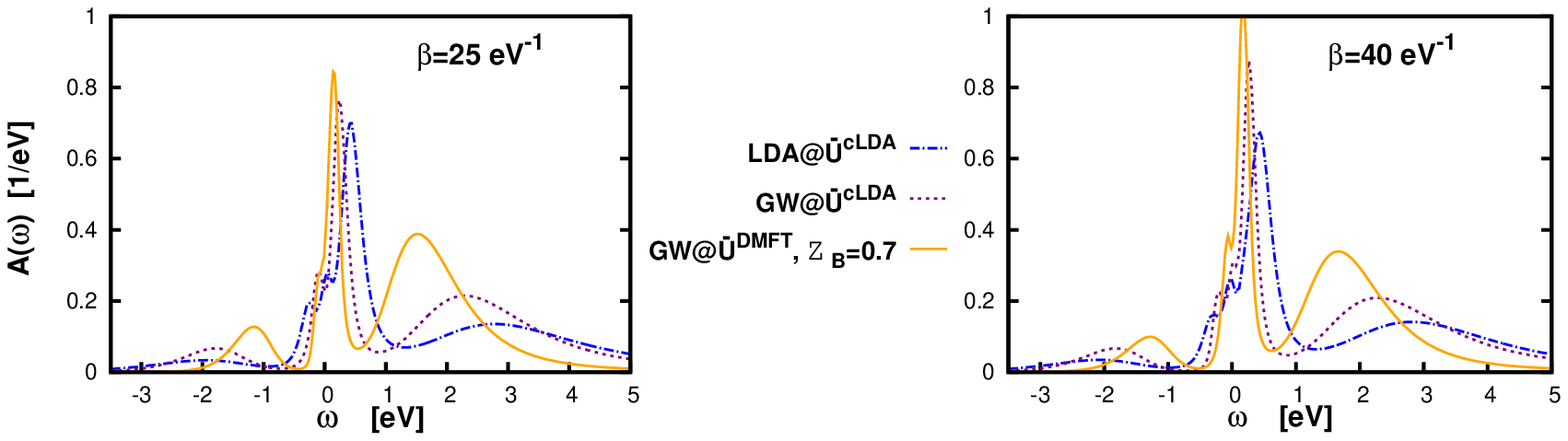}
\includegraphics[width=16cm]{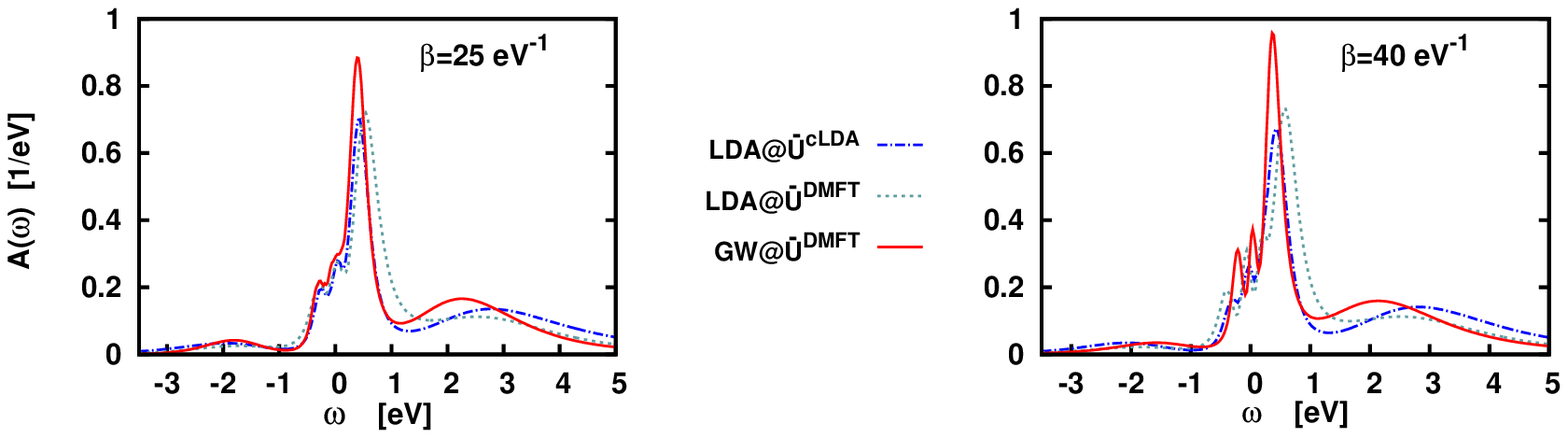}
\end{center}
\caption{(Color online) Spectral function for SrVO$_3$ computed in five different ways as in Fig.\ \protect  \ref{senergy}. At lower temperatures the central peak gets only slightly sharper and higher, 
although the temperature effects from $\beta=25$ to 40\,eV$^{-1}$ are small.}
\label{spectra}
\end{figure*}

Next, we compare the ${\mathbf k}$-integrated spectrum in
Fig.\ \ref{spectra}. At low-frequency we find the same trends as for the self-energy results: The ``standard'' $GW$+DMFT and LDA+DMFT at $\bar{U}^{\rm DMFT}$ and  $\bar{U}^{\rm cLDA}$, respectively,
yield a rather similar spectrum.
In particular the quasiparticle peak has a similar weight and shape.
However, a difference that we will come back to later on is found at larger frequencies: The $GW$+DMFT Hubbard bands are closer to the Fermi level in comparison to LDA+DMFT.
If we perform $GW$+DMFT and LDA+DMFT at the ``wrong'' interaction strength
(i.e., $\bar{U}^{\rm cLDA}$ and $\bar{U}^{\rm DMFT}$, respectively), we obtain
a noticeable stronger and weaker correlated solution.
 This trend is also reflected in the double occupations presented in   Table \ref{table1}.
Finally, as in the case of the self-energy, the $GW$+DMFT$@\bar{U}^{\rm DMFT}$,${\cal Z}_B\!=\!0.7$ solution is much more strongly correlated, with  Hubbard side bands at much lower energies. 

\section{Comparison to photoemission spectroscopy}
\label{Sec:experiment}
An obvious question is whether LDA+DMFT or $GW$+DMFT yields
``better'' results. This question is difficult to answer
and for the time being we resort to a comparison with experimental 
photoemission  spectroscopy (PES)\cite{SrVO3exp}. However, one should be well
aware of the limitations of such a comparison. On the theory side,
the involved approximations common to
the calculations, as e.g. neglecting non-local correlations beyond the
DMFT and GW level, or further effects, such as the electron-phonon coupling
or the photoemission matrix elements,
might bias the theoretical result in one way or  the other.
On the experimental side, care is in place, as well. The PES
results  considerably
improved in the last years due to better
photon sources. Furthermore, in Ref.\ \onlinecite{SrVO3exp} an oxygen $p$-background has been
subtracted, which by construction removes all
spectral weight below the region identified as the lower Hubbard band.

\begin{figure}

\includegraphics[width=8cm]{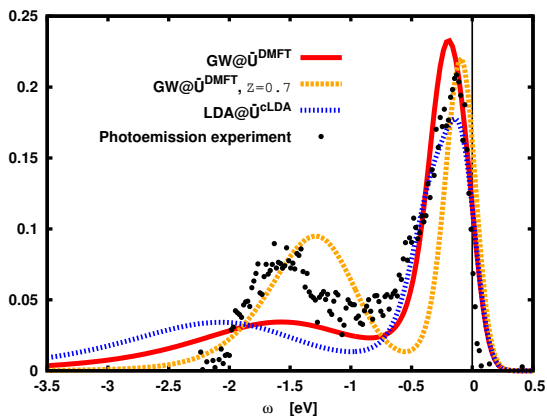}
\caption{(Color online) Comparison of LDA+DMFT$@U^{\rm cLDA}$,  $GW$+DMFT$@U^{\rm DMFT}$ (without and with Bose renormalization ${\cal Z}_B\!=\!0.7$) and experiment. The position of the lower Hubbard band is better reproduced in  $GW$+DMFT whereas the central peak is similar in  LDA+DMFT  and $GW$+DMFT. The Bose renormalization $GW$+DMFT differs considerably (photoemission spectra reproduced from Ref.\ \onlinecite{SrVO3exp}).}
\label{compexp}
\end{figure}

Fig.\ \ref{compexp} compares the proposed LDA+DMFT and $GW$+DMFT (with and without Bose renormalization) with PES experiment. To this end, the theoretical results have been multiplied with the Fermi function at the experimental temperature of $20$K and broadened by the
experimental resolution of 0.1 eV. The height of the PES spectrum has been
fixed so that its integral yields 1, i.e., accommodates 1 $t_{2g}$-electron, as in theory.

The $GW$+DMFT$@\bar{U}^{\rm DMFT}$ and LDA+DMFT$@\bar{U}^{\rm cLDA}$ have a quite similar quasiparticle peak, which also well agrees with experiment, as it was already indicated by
the quasiparticle renormalization factor. A noteworthy difference is the position
of the lower Hubbard band which is at  $-2$\,eV for LDA+DMFT$@\bar{U}^{\rm cLDA}$ and
$\sim -1.6$\,eV for $GW$+DMFT$@\bar{U}^{\rm DMFT}$. The latter is 
in agreement with experiment and a result of the reduced $GW$ band width. Let us note that the
 sharpness and height of the lower Hubbard
band very much depends on the maximum entropy method, which tends to overestimate the broadening of the high-energy spectral features. Hence, only the  position
and weight is a reliable result of the calculation.

As we have already seen, the Bose-factor renormalized $GW$+DMFT$@\bar{U}^{\rm DMFT}$,${\cal Z}_B\!=\!0.7$ calculation is  distinct from both, $GW$+DMFT$@\bar{U}^{\rm DMFT}$ and LDA+DMFT$@\bar{U}^{\rm cLDA}$. It is also different from experiment with a much more narrow
quasiparticle peak and a lower Hubbard band much closer to the Fermi
level. A similar difference between static $U$ on the one side and  frequency dependent $U$  was reported in
 Ref.\ \onlinecite{P7:Casula12b}. A difference of this magnitude is
hence to be expected. In the course of finalizing this paper, we became aware of  Ref.\ \onlinecite{Tomczak12}, in which Tomczak {\em et al.\ }
report a $GW$+DMFT calculation with the full frequency dependence of the cRPA
interaction for SrVO$_3$
obtaining good agreement with experiment as well.

\section{Conclusion}
\label{Sec:conclusion}
We have carried out a careful comparison of LDA+DMFT, $G_0W_0$+DMFT and experiment for the case of SrVO$_3$,
which is often considered to be a ``benchmark'' material for new methods.
To this end, the LDA or $G_0W_0$ quasiparticle bandstructure was projected onto maximally localized Wannier orbitals for the $t_{2g}$ bands. For these in turn correlation effects
have been calculated on the DMFT level.
If we take the locally unscreened RPA interaction (or the similar cRPA one) for the  $GW$+DMFT and the cLDA interaction for LDA+DMFT,  the two approaches yield rather similar self energies and spectral functions at the Fermi level. These also agree rather well with photoemission spectroscopy. A noteworthy difference between these two calculation is found, however, for the position of the lower Hubbard band,
 which is better reproduced in $GW$+DMFT. Similar spectra were also obtained
by Tomczak {\em et al.} \cite{Tomczak12} using a $GW$+DMFT calculation including the frequency dependence of the interaction.

 We thank R. Arita, S. Biermann, K. Nakamura, Y. Nomura, M. Imada, S. Sakai,
 and J. Tomzcak for many helful discussions.
We acknowledge  financial support from  Austrian Science Fund (FWF)
through project ID I597 (as part of the DFG Research Unit FOR 1346)
and the  SFB ViCom F4103. Calculations have been done on the Vienna Scientific Cluster (VSC).



\begin{thebibliography} {}
\bibitem{NoteUDMFT} Note that the values differ slightly from the VASP values publsihed in Ref.\ \onlinecite{P7:Nomura12} since with additional data,
we have further improved the extrapolation.
\bibitem{notekinks}
Note that due to the presence of kinks there are actually two such
renormalization factors \cite{Nekrasov05b,Byczuk06a,Kinkexp}: a Fermi liquid $Z_{\rm FL}$
for the renormalization at the lowest energies and a second $Z_{\rm CP}$ for higher energies. 
The latter also corresponds to the overall weight of the central peak.
With the energy resolution in Fig.\ \ref{senergy}
being limited by the discrete Matsubara frequencies 
the $Z$ of  Table \ref{table1} still rather corresponds to $Z_{\rm CP}$ as do
the experimental valuies of \onlinecite{SrVO3exp,Zexp}.
\end{thebibliography}
\end{document}